\documentclass[a4paper]{article}

\usepackage{amsmath}
\usepackage{amsthm}
\usepackage{amssymb}
\usepackage{cite}
\usepackage{graphicx}

\usepackage{authblk}

\newcommand{\cbar}{\mbox{$\bar c$}}
\newcommand{\ccbar}{\mbox{$c \bar c$}}
\newcommand{\dsbar}{\mbox{$\bar D^\ast$}}
\newcommand{\dsz}{\mbox{$\bar D^{\ast 0}$}}
\newcommand{\dsm}{\mbox{$D^{\ast -}$}}
\def\Dbar{\bar{{D}}{}}
\def\rt#1^^I{\sqrt{#1}}

\begin{document}
%
%
%
%
\title{X(3872) as a hybrid state of the charmonium and the hadronic molecule}

\author[1]{Makoto Takizawa\thanks{takizawa@ac.shoyaku.ac.jp}}
\author[2]{Sachiko Takeuchi\thanks{s.takeuchi@jcsw.ac.jp}}
\affil[1]{Showa Pharmaceutical University, Machida, Tokyo, 194-8543, Japan}
\affil[2]{Japan College of Social Work, Kiyose, Tokyo, 204-8555, Japan}

\renewcommand\Authands{ and }
%

%
\maketitle
\begin{abstract}%
In order to understand the structure of the $X$(3872), 
the $c \bar c$ charmonium core state which couples
to the $D^0 \bar D^{\ast 0}$ and 
$D^+ D^{\ast -}$ molecular states is studied.
The strengths of the couplings between the charmonium state and 
the hadronic molecular states are determined so as to reproduce the 
observed mass of the the $X$(3872).
The attraction between $D$ and $\dsbar$ is determined 
so as to be consistent with the observed
$Z_b^{\pm,0}$(10610)
and $Z_b^{\pm,0}$(10650) masses.
The isospin symmetry breaking is introduced by the mass differences
of the neutral and the charged $D$ mesons.  
The structure of the $X(3872)$ we have obtained is not just a
 $D^0 \bar D^{\ast 0}$ hadronic molecule 
but the charmonium-hadronic molecule hybrid state. 
It consists of about 6\%  $\ccbar$ charmonium,
69\%  isoscalar $D \bar D^\ast$ molecule and 26\% 
isovector $D \bar D^\ast$ molecule. 
This explains many of 
the observed properties of 
the $X$(3872), such as the isospin symmetry breaking, 
the production rate in the $p \bar p$ collision, 
a lack of the existence of the $\chi_{c1}(2P)$ peak
 predicted by the quark model,
and the absence of the charged $X$.
The same picture can be applied to other heavy two-meson $S$-wave 
systems,
where the states predicted by the quark model are not observed above the 
thresholds.
\end{abstract}


\section{Introduction \label{int}}
The $X$(3872) state was first observed in 2003 by Belle in 
$B^{\pm} \to J/\psi \, \pi^+ \, \pi^- \, K^{\pm}$ \cite{Belle03} and was confirmed by
CDF \cite{CDF04}, D0 \cite{D004} and $BABAR$ \cite{BaBar05} collaborations. 
The observed masses of the $X$(3872) in the $J/\psi \, \pi^+ \, \pi^-$ channel 
from the recent measurements of the charged and the
neutral B decays are 
$(3871.4 \pm 0.6 \pm 0.1)$ MeV and  $(3868.7 \pm 1.5 \pm 0.4)$ MeV,
respectively \cite{BaBar08}. Those from the $p \bar p$ and the $p p$ collisions are
$(3871.61 \pm 0.16 \pm 0.19)$ MeV \cite{CDF09} and 
$(3871.95 \pm 0.48 \pm 0.12)$ MeV \cite{LHCb12}.
The average mass given by the particle data group in 2012 \cite{PDG12} is 
$(3871.68 \pm 0.17)$ MeV, which is 0.16 MeV below the $D^0 \dsz$ threshold.
The full width is less than 1.2 MeV.

As for the spin-parity quantum numbers of the $X$(3872), the angular distributions and correlations
of the $\pi^+ \pi^- J/\psi$ final state have been studied by CDF \cite{CDF07} and 
they concluded that the pion pairs originate from $\rho^0$ mesons and that
the favored quantum numbers of the  $X$(3872) are $J^{PC} = 1^{++}$ and $2^{-+}$.
Recent analyses support $J^{PC} = 1^{++}$ interpretation \cite{Hanhart12}.
$BABAR$ has found the evidence of the radiative decays of $X(3872) \to \gamma J/\psi$ 
with 3.4-3.6 $\sigma$ significance \cite{BaBar06,BaBar09}, 
which implies that the  $C$-parity of $X$(3872) is positive.
Though we assume that $X$(3872) is $J^{PC} = 1^{++}$,
whether the quantum number is $1^{++}$ or $2^{-+}$ is still 
an issue of the discussion and more experimental data are certainly necessary.

Since the first observation of the $X$(3872), 
it has received much attention because its features are difficult to explain
if a simple $c \bar c$ bound state of the quark potential model is assumed \cite{Barnes04}.
$X(3872)$ is one of the promising candidates of the exotic states reviewed in Ref.\
\cite{Swanson06,Godfrey08,Brambilla11}.
Many kinds of structures  have been suggested for the $X$(3872) from the theoretical side, 
such as a tetraquark structure 
\cite{Maiani05,Ebert06,Terasaki07,Dubnicks10}, 
a $D^0 \dsz$ molecule \cite{Close04,Voloshin04,Swanson04,Tornqvist04,AlFiky06,Ding09,Lee09,Gamermann10} 
and a charmonium-molecule hybrid \cite{Matheus09,Danilkin10,Coito11}.
We also employ this hybrid picture and argue that that is most appropriate.

One of the important properties of the $X$(3872) is its isospin structure.
The branching fractions measured by Belle \cite{Belle05} is
\begin{equation}
\label{TT_eq:1_1}
  \frac{Br(X \rightarrow \pi^+ \pi^- \pi^0 J/\psi)}{Br(X \rightarrow \pi^+ \pi^- J/\psi)} 
  = 1.0 \pm 0.4 \pm 0.3 \, ,
\end{equation}
and $(0.8 \pm 0.3)$ by $BABAR$ \cite{BaBar10}. 
Here the two-pion mode originates from the isovector $\rho$ meson 
while the three-pion mode comes from the isoscalar $\omega$ meson. 
So, the eq.\ (\ref{TT_eq:1_1}) indicates strong isospin violation. 
M.\ Suzuki has estimated the kinematical suppression factor 
including the difference of the vector meson decay width and 
obtained the production amplitude ratio \cite{Suzuki05} using Belle's value
\begin{equation}
\label{TT_eq:1_2}
  \left| \frac{A(\rho J/\psi)}{A(\omega J/\psi)} \right|
  = 0.27 \pm 0.02 \, .
\end{equation}
Usual size of the isospin symmetry breaking is at most a few \%.
It is interesting to know what is the origin of this strong isospin symmetry breaking.
In \cite{Gamermann09}, this problem has been studied by using the chiral unitary model and
the effect of the $\rho$-$\omega$ mixing has been discussed in \cite{Terasaki10}.
It was reported that
both of the approaches can explain the observed ratio given in Eq.~(\ref{TT_eq:1_1}),
but at present, consensus on the mechanism of the large isospin symmetry breaking has yet to be reached.
We will show in this work that the mass difference of the $D^0\dsz$ and the $D^+D^{*-}$ 
thresholds gives enough amount of the isospin violation to explain the experiments.

The production processes have been
studied in \cite{Braaten04,Braaten05,Bignamini09,Zanetti11}.
It seems that a pure molecule picture 
cannot explain the production rate of the $X$(3872) 
in the $p\bar p$ collision well \cite{Bignamini09}. 
There, the production rate of the $X$(3872) 
 is about 1/20 of the rate of $\psi (2S)$, which 
suggests that $X$(3872) has to have, 
by a very rough estimate, the order of 5\% of the 
 $c\cbar$ component.

The hadronic decays of the $X$(3872) are investigated
in \cite{Swanson04-2,Braaten05-2,Navarra06,Braaten07,Dubynskiy08,Braaten08,Fleming08,Artoisenet10,Fleming12}.
As for the radiative decays,
as seen in \cite{Colangelo07,Dong08,Nielsen10,Wang11,Harada11,Mehen11,Dubnicka11,Ke11,Badalian12},
the existence of the core seems to be required, but the results depend on details of the wave function. 
Here we assume that the $c\cbar$ core 
is created by the weak interaction in the $B$ decay as $B\rightarrow (c\cbar)+K$,
and investigate the transfer strength from the $c\cbar$ to $D\dsbar$.
We only investigate the hadronic mode
and will discuss the radiative decay elsewhere.

The $X$(3872) exists above the open charm threshold, $D\Dbar$.
Below this threshold, the $c \cbar$ mass spectrum is well predicted
by a simple quark model.
The model, however, failed to predict the masses above the open charm
threshold for the $D\Dbar$ or $B\bar B$ $S$-wave sector.
In this work, we also show that the $c\cbar$ peak above the threshold can actually disappear
by introducing the $c\cbar$-$D\dsbar$  coupling.

It is also an important issue that whether the charged partner of the $X$(3872) exists 
as a measurable peak or not.
$BABAR$ has searched such a state in the $X \to \pi^- \pi^0 J/\psi$ 
channel and found no signal \cite{BaBar05-2}. 
The hybrid picture, where the coupling to the $c\cbar$ core is
essential to bound the neutral $X$, is consistent with the
absence of the charged $X$.

Recently, $Z_b(10610)^{\pm,0}$ and $Z_b(10650)^{\pm,0}$ ($J^P$=$1^+$)
 resonances 
have been found in the $\Upsilon(5S)$ decay to $\Upsilon(nS)\pi^+\pi^-$ 
($n$=1,2,3)
and $h_b(mP)\pi^+\pi^-$ ($m$=1,2)
reactions\cite{Belle12}.
The masses of these resonances are just above the
$B\bar B^*$ and the $B^*\bar B^*$ thresholds, respectively;
the main component 
is considered to be  the $B^{(*)}\bar B^*$ two-meson state.
This means that there exists an almost-zero-energy bound state (or resonance)
in each of the  $D\bar D^*$ and the $B^{(*)}\bar B^*$ systems.
In order to make such states, the attraction in the $D\bar D^*$ system
 is considered to be about 2.7 times as strong as that of 
the $B^{(*)}\bar B^*$ system because
the reduced mass of the  $D\bar D^*$ system, 967 MeV, 
is about 2.7 times
as light as that of the $B^{(*)}\bar B^*$ system, 2651 MeV.
On the other hand,
the interaction
between the $D$ and $\bar D^*$ mesons is probably
about the same size as that between  the $B^{(*)}$ and $\bar B^*$ mesons.
We argue that the extra attraction 
required for the $X(3872)$ comes, at least mainly, from its
coupling to the $c\cbar$ core,
which is absent in these isovector $Z_b$ systems.

In this article, we present a hybrid picture where $X$(3872)
is $J^{PC}=1^{++}$ and  consists of 
$D^0\dsz$, $D^+D^{*-}$, and the $2\,{}^3\!P_1$ $c\cbar$ core,
which stands for the $\chi_{c1}(2P)$ if observed.
A separable $D\dsbar$ interaction is introduced,
whose strength is determined so as to give a zero-energy bound state
when it applied to the $B^{(*)}\bar B^*$ systems.
The rest of the required attraction to form the $X$(3872) are assumed to come from 
the $c\cbar$-$D\dsbar$ coupling.
The coupling strength is determined so as to give the observed $X$(3872) mass.
The $c\cbar$ core mass is taken from the quark model result,
and the cutoff is chosen by considering the $c\cbar$ core size.
As we will discuss later, the behaviors of the $X$(3872)
do not depend strongly on the detail of the interactions.
Main parameters of the present model are the overall strength of the
two-meson interaction and that of the coupling,
which are essentially determined from the masses of $X$(3872) and $Z_b$'s.
This simple picture, however, 
is found to be consistent with many of the experiments, 
such as the isospin symmetry breaking
in the $X$(3872) decay, the production rate of $X$(3872) in the $p\bar p$ collision,
and the absence of $\chi_{c1}(2P)$ peak or the charged $X$,
in addition to the mass of $X$(3872) and $Z_b$'s, which are the inputs.

It should be noted here that the quark number is not the conserved quantity in 
QCD and our treatment of taking the $\ccbar$ and $D\dsbar$ as the orthogonal 
states is an approximate one. In the low-energy QCD, the spontaneous chiral 
symmetry breaking occurs and the light quarks get the dynamical masses.
In such a situation, the treatment of taking the $\ccbar$ and $\ccbar u \bar u$
($\ccbar d \bar d$) as the orthogonal states seems to be acceptable, since 
these two states are energetically different. 
In order to study the structure of the exotic hadrons, 
how to count the quark number is an issue of the discussions. 
Three methods have been proposed to observe the number of the valence quarks 
in the hadron.
The first one is to measure the elliptic flow 
in relativistic heavy ion collisions \cite{Nonaka04} while the second one is 
to measure the nuclear modification ratios in heavy ion collisions \cite{Maiani07}.
The last one is to use the fragmentation functions \cite{Hirai08}.
We hope some of these methods will be applied to the $X$(3872) and the quark 
component of the $X$(3872) will be determined experimentally.

Let us briefly mention features of our work in comparison to those
that also employ the charmonium-molecule hybrid model \cite{Matheus09,Danilkin10,Coito11}.
In Ref.\ \cite{Matheus09}, the hybrid structure of the X(3872) has been studied 
in the QCD sum rule approach by considering a mixed charmonium-molecular current.
They found a very deeply bound $X(3872)$,
 $m_X = (3.77 \pm 0.18)$ GeV, 97\% of whose component is a charmonium.
As we shall show in Sec.\ref{form}, the structure of the $X(3872)$ certainly depends on the 
binding energy strongly. 
In Ref.\ \cite{Danilkin10,Coito11}, the effective hadronic models with 
the charmonium-$D \bar D^{\ast}$ molecule transition interaction have been used 
as well as in the present work. 
Danilkin and Simonov  have
studied the $D \bar D^{\ast}$ production spectrum \cite{Danilkin10}.  
They obtained the strength of  the $c\bar c$-$D\bar D^*$ coupling
from the heavy quarkonium decay calculated by a quark model with a small adjustment.
They certainly found a steep rise near the threshold.
We have studied the $B \to X(3872) \, K$ or $D\bar D^*\,K$
weak decay spectrum at almost same time independently in the very similar approach 
in \cite{TT10-1,TT10-2}.  
Here, we also introduce the interaction between $D$ and $\bar D^*$,
and look into the features of the shallowly bound $X(3872)$.
The work in Ref.\ \cite{Coito11} they examined the 
effects of the Okubo-Zweig-Iizuka forbidden $\rho^0 J/\psi$ and $\omega J/\psi$ channels.
Since the main decay modes of the $X(3872)$ are the $X(3872) \to \pi^+ \pi^- J/\psi$ and 
$X(3872) \to \pi^+ \pi^- \pi^0 J/\psi$, 
their inclusion is certainly important.
As shown in \cite{Coito11}, however,
this effect on the pole position seems rather small.
To avoid the complication we discuss it elsewhere.
In the present study, we have introduced the attractive interaction between
$D$ and $\bar D^\ast$ mesons with the coupling strength being consistent with 
the observed $Z_b (10610)$ and $Z_b (10650)$ masses. 
This point is new to the previous two studies and we consider that 
we can successfully draw the consistent picture of the observed
exotic hadrons $X(3872)$, $Z_b (10610)$ and $Z_b (10650)$.

One of the authors (S. T.) has studied the $X$(3872) using a quark potential model by introducing 
an extra ($q \bar q$) pair to a $c \bar c$ system \cite{Takeuchi07} and found a shallow bound state of 
$q \bar q c \bar c$ with $J^{PC} = 1^{++}$.
Recently, an elaborate study has been done in the quark potential model \cite{Ortega10}.
They have performed the coupled channel calculations including two and four-quark configurations
using the $^3P_0$ model and found a good agreement with the experimental data.
The purpose of the present work is to make the situation of the $X$(3872) clearer by studying 
the role of the $c \bar c$ core state, which couples to the 
$D^0 \dsz$ and $D^+ \dsm$ molecular states, with a simple hadronic model.
This approach will complement the picture given by the quark model approach.

This paper is organized as follows.
In Sec.\ref{form}, the calculation of the  $X$(3872) state
is given. In Sec.\ref{spectra}, we discuss the transition strength of the
weak decay of B meson: $B \to X(3872) K$ or $D\bar D^*\,K$
 using the Green's function approach. 
In Sec.\ref{DDint}, we study the effects of the interaction between 
the $D$ and \dsbar\ mesons.
We discuss the possibility of the other exotic hadrons by the 
present mechanism in Sec.\ref{ExoticHadrons}.
Finally, Sec.\ref{conc} is devoted to summary of this paper.
\section{$X$(3872)}
\label{form}%
\begin{figure}[!tb]
\centering
\includegraphics[width=0.6\columnwidth]{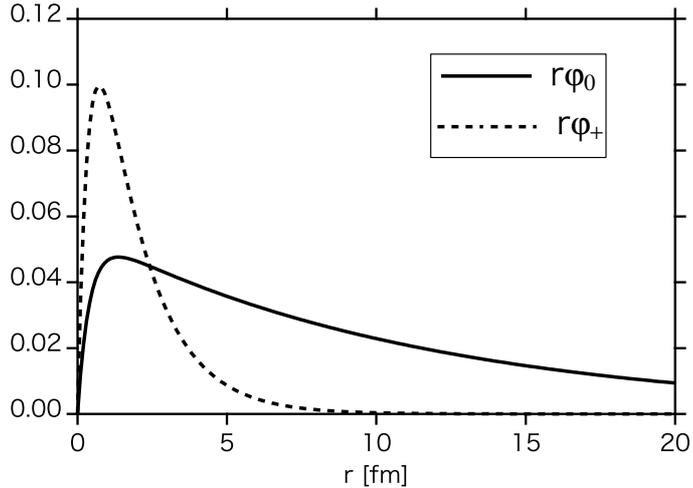}
\caption{The $D\bar D^*$ components of the $X$(3872) wave function
for the  $m_X = 3.87168$ GeV  and $\Lambda=0.5$ GeV case.
$D^0 \dsz$ wave function, $r \varphi(r)_0$,  is plotted by the solid line,
 and
$D^+ \dsm$ wave function, $r \varphi(r)_+$, by the dashed line. 
}
\label{TT_fig:2_1}       \end{figure}

We argue that the $X$(3872) state is a superposition of the 
$c \bar c$ core state, the $D^0 \dsz$ hadronic molecular
state, and the $D^+ \dsm$ hadronic molecular states.
So, the wave function of the $X$(3872) in the center of the mass frame is represented by 
\begin{equation}
\label{TT_eq:2_1}
  | X \rangle = c_1 \, | c \bar c \rangle + c_2 \, | D^0 \dsz \rangle 
  + c_3 \, | D^+ \dsm \rangle \, .
\end{equation}
The $D^0 \dsz$ and $D^+ \dsm$ molecular states are given by
\begin{align}
\label{TT_eq:2_2}
  | D^0 \dsz \rangle & = 
    \int d^3 {\pmb q} \, \varphi_0({\pmb q}) | D^0 \dsz ({\pmb q}) \rangle
 \, ,
 \\
\label{TT_eq:2_3}
  | D^+ \dsm \rangle & = 
    \int d^3 {\pmb q} \, \varphi_+({\pmb q}) | D^+ \dsm ({\pmb q}) \rangle
 \, ,
\end{align}
where ${\pmb q}$ represents the relative momentum of the 
$D$ and $\dsbar$ mesons.
The normalization of the states are
\begin{equation}
\label{TT_eq:2_4}
\begin{split}
  \langle D^0 \dsz ({\pmb q'})| D^0 \dsz ({\pmb q}) \rangle & = 
  \langle  D^+ \dsm ({\pmb q'}) | D^+ \dsm ({\pmb q}) \rangle \\
   & = \delta^3({\pmb q'} - {\pmb q})
 \, .
\end{split}
\end{equation}
Here $\varphi_0({\pmb q})$ and $\varphi_+({\pmb q})$ are the momentum representation of 
the wave functions of the $D^0 \dsz$ and $D^+ \dsm$ hadronic molecular states, respectively.
The charge conjugation is assumed to be positive 
throughout this paper.
We assume these three states ($| c \bar c \rangle$, $| D^0 \dsz \rangle$
and $| D^+ \dsm \rangle$) are the orthonormal states.
If $| D^0 \dsz \rangle$ and $| D^+ \dsm \rangle$ are the spatially wide objects,
this assumption seems to be reasonable. As we shall show in Fig.~\ref{TT_fig:2_1},
indeed $| D^0 \dsz \rangle$ and $| D^+ \dsm \rangle$ are the wide objects.

We introduce a coupling between
the $c \bar c$ core state and the $D \dsbar$ states in the isospin symmetric manner.
Since we are looking into the low energy region, the results do not depend much 
on the shape of the interaction.
Thus, we take a monopole-type coupling as:
\begin{equation}
\label{TT_eq:2_5}
\begin{split}
  \langle D^0 \dsz ({\pmb q})| V | c \bar c \rangle & = 
  \langle  D^+ \dsm ({\pmb q}) | V | c \bar c \rangle \\
  & = \frac{g}{\sqrt{\Lambda}} \left( \frac{\Lambda^2}{{q}^2 + \Lambda^2} \right)
 \, .
\end{split}
\end{equation}
The interaction we have introduced above 
causes effectively an attraction 
for the $X$(3872) because its energy is lower than the mass of the $c\cbar$ core, $m_{c\bar c}$.
In this section, we 
ignore the direct interactions between the $D$ and $\bar D^*$ mesons;  
as we will discuss later in Sec.\ref{DDint}, 
the coupling to the $c\cbar$ core seems more important to make the $X$(3872) than 
the direct $D\dsbar$ attraction.

Here we consider only the relative $S$-wave states of these two mesons
in the non-relativistic scheme because 
the $X$(3872) is close to the threshold.
The Schr\"odinger equation to solve is 
\begin{equation}
\label{TT_eq:2_6}
  \begin{pmatrix}
  m_{c \bar c} - E & V & V \\
  V & m_{D^0} + m_{D^{\ast 0}} + \frac{{\hat p}^2}{2 \mu_0} - E & 0 \\
  V & 0  & m_{D^+} + m_{D^{\ast -}} + \frac{{\hat p}^2}{2 \mu_+} - E
  \end{pmatrix}
   \,
  \begin{pmatrix}
  c_1 \, | c \bar c \rangle \\
  c_2 \, | D^0 \dsz \rangle \\
  c_3 \, | D^+ \dsm \rangle
  \end{pmatrix} \,  = \,
  \begin{pmatrix}
  0 \\
  0 \\
  0
  \end{pmatrix} \,    
 \, ,
\end{equation}
with
\begin{equation}
\label{TT_eq:2_7}
 \frac{1}{\mu_0} = \frac{1}{m_{D^0}} + \frac{1}{m_{D^{\ast 0}}}
 \, , \qquad
 \frac{1}{\mu_+} = \frac{1}{m_{D^+}} + \frac{1}{m_{D^{\ast -}}}
 \, .
\end{equation}

Since the interaction we employ is separable, we can solve this Schr\"odinger 
equation analytically.
The bound state energy is obtained by solving the following equation.
\begin{equation}
\label{TT_eq:2_8}
 m_{c \bar c} - E - F_0(E) -F_+(E) = 0
 \, ,
\end{equation}
with
\begin{equation}
\label{TT_eq:2_9}
 F_0(E) = \int  
 \frac{d^3 {\pmb q}}{\left( m_{D^0} + m_{D^{\ast 0}} + \frac{{ q}^2}{2 \mu_0} \right) - E}
 \left( \frac{g \, \Lambda^{3/2}}{{q}^2 + \Lambda^2} \right)^2
 \, ,
\end{equation}
and
\begin{equation}
\label{TT_eq:2_10}
 F_+(E) = \int
 \frac{d^3 {\pmb q}}{\left( m_{D^+} + m_{D^{\ast -}} + \frac{{ q}^2}{2 \mu_+} \right) - E}
 \left( \frac{g \, \Lambda^{3/2}}{{q}^2 + \Lambda^2} \right)^2
 \, .
\end{equation}

For later convenience, we define $\alpha_0$ and $\alpha_+$ by
\begin{equation}
\label{TT_eq:2_11}
 \frac{\alpha_0^2}{2 \mu_0} = m_{D^0} + m_{D^{\ast 0}} -m_X
 \, ,
\end{equation}
and
\begin{equation}
\label{TT_eq:2_12}
 \frac{\alpha_+^2}{2 \mu_+} = m_{D^+} + m_{D^{\ast -}} -m_X
 \, ,
\end{equation}
where $m_X$ represents the observed mass of the $X$(3872).

In order to obtain the numerical results,
 we use the $D$ meson masses 
given in the 2012 Review of Particle Physics \cite{PDG12} (Table \ref{tbl:masses}).
Since we have introduced the isospin symmetric interaction $V$ 
in Eq.~(\ref{TT_eq:2_5}), the only origin of the isospin violation in the present model
is the mass difference between the charged and neutral $D$ and $D^\ast$ mesons.
\begin{table}[!b]
\caption{Meson masses and the thresholds. All the entries are in GeV.}
\label{tbl:masses}
\centering
\begin{tabular}{cccccccc}
\hline\hline
$m_{D^0}$ & $m_{D^+}$ &$m_{D^{\ast 0}}$ & $m_{D^{\ast -}}$
& $m_{D^0}+m_{D^{\ast 0}}$ &$m_{D^+}+m_{D^{\ast -}}$\\
\hline
1.86486 & 1.86962 & 2.00698 & 2.01028&3.87184&3.87990\\
\hline\hline
\end{tabular}
\end{table}%

As for the $c \bar c$ core state, 
we consider that it corresponds to the $J^{PC} = 1^{++}$ charmonium
state with the mass $m_{c \bar c}= 3.950$ GeV, the closest $c\cbar$ core to $X$.
This value is taken from the Godfrey and Isgur's results of the 
quark potential model calculation for the 2 $^3$P$_1$ $c \bar c$ state \cite{GI85}.

In the following, we will show the results as well as their dependence on 
the various assumptions.

There are two 
free parameters in the present model:  the cutoff $\Lambda$ and 
the dimensionless coupling constant $g$. 
We take typical hadron sizes for $\Lambda$: {\it e.g.},
$\Lambda = 0.3$ GeV, 0.5 GeV and 1.0 GeV. 
Then, for a given $\Lambda$, 
the coupling constant $g$ is determined so that the model reproduces the observed mass of the $X$(3872),
namely, 3.87168 GeV.
The results are given in Table \ref{TT_tab:1}.
\begin{table}[!b]
\caption{%
The values of the dimensionless coupling constant $g$
for each value of the cutoff $\Lambda$ in units of GeV.
The mass of X(3872) is $m_X = 3.87168$ GeV.
}
\label{TT_tab:1}
\centering
\begin{tabular}{c|lll}
\hline\hline
$\Lambda$ [GeV] & 0.3 & 0.5 & 1.0 \\
\hline
$g$ & 0.05435 & 0.05110 & 0.04835 \\
\hline\hline
\end{tabular}
\end{table}
\begin{table}[!b]
\caption{Coefficients of the $X(3872)$ wave function.}
\label{tbl:ccc1}
\centering
\begin{tabular}{ccccccccc} \hline\hline
$\Lambda$ & $c_1$ &$c_2$ &$c_3$& $c_{I=0}$ &$c_{I=1}$&$m_X$\\
\hline
0.3 &0.227 &$-0.947$ & $-0.228$ & $-0.831$ & $-0.508$\\ 
0.5 &0.293 &$-0.920$ & $-0.259$ & $-0.834$ & $-0.468$&3.87168\\ 
1.0 &0.404 &$-0.871$ & $-0.280$ & $-0.814$ & $-0.418$\\ 
\hline
0.5 &0.522 &$-0.727$ & $-0.447$ & $-0.830$ & $-0.198$&3.8687\phantom{0}\\ 
\hline\hline
\end{tabular}
\end{table}%
The wave function we have obtained: 
\begin{align}
\label{TT_eq:2_13}
 | X \rangle  =& c_1 \, | c \bar c \rangle+c_2 \, | D^0 \dsz \rangle 
    +c_3 \, | D^+ \dsm \rangle  \notag \\
                =& c_1 \, | c \bar c \rangle +c_{I=0} \, | D \dsbar ; I=0 \rangle 
  +c_{I=1} \, | D \dsbar ; I = 1 \rangle 
 \, .
\end{align}
The values of $c$'s are shown in Table \ref{tbl:ccc1} for each of the cutoff values.
It seems that the overall feature of the  admixture of each component does not depend much 
on the value of $\Lambda$,
which is not surprising because 
a very shallow state does not depend much on detail of the potential.
The main component of the $X$(3872) state is $| D^0 \dsz \rangle$,
reflecting the fact that the mass of the $X$(3872) is only $0.16$ MeV below the 
$D^0 \dsz$ threshold.
The amplitude of the  $| D^+ \dsm \rangle$ component is much smaller 
because the $D^+ \dsm$ threshold is 
$8.22$ MeV above the mass of the $X$(3872).
The size of the isospin symmetry breaking we have obtained 
for the averaged mass, 3.87168 GeV, seems to be roughly consistent
with the one estimated from the experiments given by Eq.~(\ref{TT_eq:1_2}).

Let us emphasize that we have obtained a measurable amount
 of the $| c \bar c \rangle$ component. 
In the present scheme, 
the origin of the attraction 
is the coupling between 
the $D \dsbar$ component and $c \bar c$ core state.
So, it is natural to have a certain amount of  
$| c \bar c \rangle$ component in the $X$(3872) state.

The mass of the $c\cbar$ core is taken from the quark model calculation.
It may differ because of the model assumption.
The slight change of its mass, however, does not change our results much.
For example, for $m_{{c\bar c}}=3.93$ or 3.97 GeV, which
we take because $\pm$ 20 MeV  is typical 
ambiguities of the quark models,
the $c\cbar$ component becomes
11\% and 7\%, respectively. 
There is no drastic change in the results.

We take only this state as the $c \bar c$ core state in this paper
because it has the closest mass to that of X(3872).
The mass of the 1 $^3$P$_1$ $c \bar c$ state is, for example, around 3.5 GeV and therefore 
its coupling to the $X(3872)$ will be suppressed \cite{Ortega10}. 
In our calculation which includes the lower $c\cbar$ state with the same coupling size,
the probability of the 1 $^3$P$_1$ $c \bar c$ state
is found to be about 1/20 of that of the 2 $^3$P$_1$ $c \bar c$ state. 
The existence of another core component may change the nature of the $\gamma$-decay
of $X$(3872), where a large cancellation occurs and results are very sensitive to the 
wave function \cite{Barnes:2005pb,Badalian:2012jz}. 
We, however, look into such observables elsewhere,
and concentrate on the bulk feature of $X$(3872) in this work.

The S-wave state of the $D^\ast \dsbar$ channel is able to couple with 
the $J^{PC} = 0^{++}$ charmonium state and the threshold of 
the $D^\ast \dsbar$ channel is about 140 MeV above the $X$(3872) mass. 
We, therefore, should examine whether the $D^\ast \dsbar$ channel can contribute 
to the structure of the $X$(3872). We have performed the calculation of the 
$X$(3872) structure with the $D^\ast \dsbar$ channel and the result has been
that the $D^\ast \dsbar$ component of the $X(3872)$ is about 2\%, 
reasonably small.

Experimental  uncertainty of the X(3872) mass still exists.
So, we solve the system also for $m_X = 3.8687$ GeV.
This mass is the one
determined from the neutral B meson decay data, 
and the lightest mass among the ones given by the experiments. 
Now the binding energy becomes 3.14 MeV instead of the one
corresponding to the average mass, 0.16 MeV.
The value of $g$ to form the lighter mass becomes 0.05625,
which is 1.1 times as large as that of the average mass, 0.05110.
In order to form the more deeply bound $X$,
the dimensionless coupling constant $g$ is required to
be larger.
The coefficients of the wave function are listed in Table \ref{tbl:ccc1}.
The size of the $c \cbar$-core component also becomes larger:
it changes from $(0.293)^2 \simeq 0.086$ to $(0.522)^2 \simeq 0.272$ 
as $m_X$ changes from 3.87168 to 3.8687 GeV for the case of $\Lambda = 0.5$ GeV.
The size of the $c \cbar$-core component in the $X(3872)$ is found to be sensitive
to the binding energy of the state.
The amount of the isospin symmetry breaking
depends also on the binding energy of $X$;
The symmetry breaking occurs because of 
the difference of the binding energies of the $X$(3872) from the two thresholds, i.e., 
$D^0 \dsz$ and $D^+ \dsm$. 
For $m_X = 3.87168$ and 3.8687 GeV, 
 the ratios of the size of the isovector to the isoscalar
$D \dsbar$ components are 0.315 and 
0.057, respectively.
When the mass of the $X(3872)$ becomes smaller, 
namely, the binding energy becomes larger, 
the effects of the threshold difference becomes smaller, 
and the isospin violation becomes smaller.

\begin{figure}[!tb]
\centering
\includegraphics[width=0.6\columnwidth]{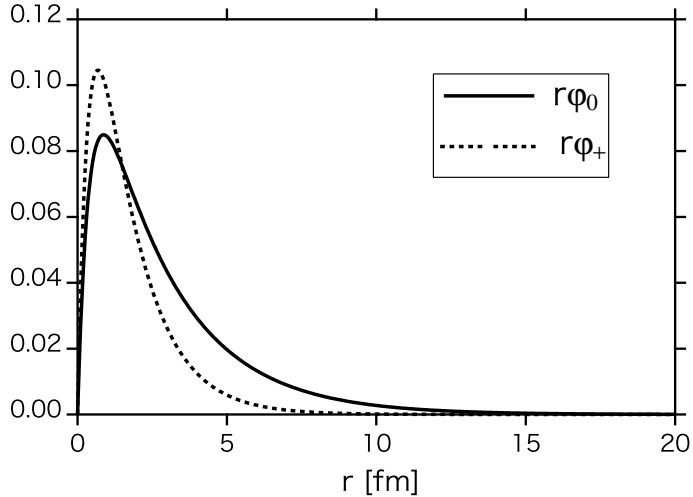}
\caption{The $D\bar D^*$ components of the $X$(3872) wave function
for the  $m_X = 8687$ GeV  and $\Lambda=0.5$ GeV case.
The legend is as for Fig.\ \ref{TT_fig:2_1}. %
}
\label{TT_fig:2_2} 
\end{figure}

Let us show the shape of the obtained wave functions.
The explicit expressions of the wave functions in the coordinate space  
are
\begin{align}
\label{TT_eq:2_18}
r \varphi(r)_0 = \left( \frac{\pi}{2} \right)^{1/2} \, \frac{N_0}{\Lambda^2 - \alpha_0^2} 
\left( e^{-\alpha_0 r} - e^{-\Lambda r} \right)
 \, ,
\end{align}
and 
\begin{align}
\label{TT_eq:2_19}
r \varphi(r)_+ = \left( \frac{\pi}{2} \right)^{1/2} \, \frac{N_+}{\Lambda^2 - \alpha_+^2} 
\left( e^{-\alpha_+ r} - e^{-\Lambda r} \right)
 \, ,
\end{align}
with 
\begin{align}
\label{TT_eq:2_20}
N_0 = 2 \mu_0 \frac{g}{\sqrt{\Lambda}} \left( \frac{c_1}{c_2} \right) \, ,  \quad
N_+ = 2 \mu_+ \frac{g}{\sqrt{\Lambda}} \left( \frac{c_1}{c_3} \right) \, .
\end{align}
Each of the neutral and the charged $D\dsbar$ components 
of the wave function of the bound state 
with $\Lambda = 0.5$ GeV and 
$m_X = 3.87168$ GeV
is shown in Fig.~\ref{TT_fig:2_1}.
It is also found that 
the radius of the $D^+ \dsm$ component is much smaller than that of  
$D^0 \dsz$. 
In Fig.~\ref{TT_fig:2_2}, we show the wave function of the $X(3872)$ 
also for $m_X = 3.8687$ GeV.
One finds that the size of the
bound state, especially the size of $D^0 \dsz$ component,
becomes much smaller than that in Fig.~\ref{TT_fig:2_1}
though it is still much larger than the usual charmonium, whose rms $\lesssim$ 1 fm \cite{GI85}.
\section{Spectrum}
\label{spectra}
In this section, we investigate the transition strength $S(E)$ of the  $B$ meson weak decay:
$B \to c\cbar K$ and $c\cbar$ to $X$(3872) or $D\dsbar$.
The $X$(3872) appears as a bound state in the spectrum.
This spectrum does not correspond directly to the observed pion distribution
in the $X(3872)\rightarrow J/\psi \pi^n$ experiments.
By looking into the $D\dsbar$ spectrum, however, one can see that 
the strength of $D\dsbar$ gathers around the threshold, 
and that the peak corresponding to the $c\cbar$ core actually disappears.

In this article, we assume that the observed X(3872) 
corresponds to a very shallow bound state. 
To have such a bound state, 
the interaction must be attractive but maybe a rather weak one.
As mentioned in 
Sect.~\ref{form}, we have fixed the strength of the $D\dsbar$-$c\cbar$ coupling, $g$, 
so as to
reproduce the observed $X$(3872) mass.
In such a situation, the $c \bar c$ core state becomes a resonance
appearing in the $D \dsbar$ continuum. 
Since no sharp resonance is observed experimentally around 3.95 GeV, 
the width of this resonance should be large.
One of the issues in this section is 
whether such a `weak' attraction can give a resonance with a large decay width.

The $S(E)$  is normalized so that the production of 
the 2$^3P_1$ 
$c \bar c$ state by the weak decay is equal to one.
The vertex of the weak decay process, $B\rightarrow c\bar c+K$,
and the probability that the $c\cbar $ is in the 2$^3P_1$ configuration 
are factorized out.
We assume that among the $c\bar c$ states
produced by the weak decay, the 2$^3P_1$ 
$c \bar c$ state plays a major role to form the $X$(3872) and the $D\dsbar$ spectrum 
up to around $E\sim$ 4 GeV because the predicted mass of the 2$^3P_1$ 
$c \bar c$ state is 3.95 GeV.
Again we use the non-relativistic scheme with the relative $S$-wave,
because the reduced mass of the system is
about 1 GeV
and we only consider here up to about 0.1 GeV above the threshold.

Then, the $S(E)$ is expressed as follows.
\begin{equation}
\label{TT_eq:3_1}
 S(E) = \frac{-1}{\pi} {\rm Im} \langle c \bar c | G(E) | c \bar c \rangle \, ,
\end{equation}
with the Green's function;
\begin{equation}
\label{TT_eq:3_2}
G(E) = \frac{1}{E - \hat H + i \varepsilon} \, .
\end{equation}
Here, $E$ represents the energy transfer and $\hat H$ is 
the full Hamiltonian of the $c \bar c$-core and 
$D \dsbar$ system.
The state $| c \bar c  \rangle$ represents the center of mass system of the 
$c \bar c$ state 
with the normalization $\langle c \bar c  | c \bar c \rangle =1$,
This normalization leads the energy sum rule
\begin{align}
\int {\rm d}E\; S(E)&=1~.
\end{align}
Using the free Green's functions and 
the interaction given in Eq.\ (\ref{TT_eq:2_5}), the Green's function is represented as follows.
\begin{equation}
\label{TT_eq:3_3}
G(E) = G_1^0 + G_1^0 V G_2^0 V G_1^0 +  G_1^0 V G_3^0 V G_1^0 + \cdots \, ,
\end{equation}
\begin{equation}
\label{TT_eq:3_4}
G_1^0(E) = \frac{1}{E - m_{c \bar c} + i \varepsilon} \, ,
\end{equation}
\begin{equation}
\label{TT_eq:3_5}
G_2^0(E) = \frac{1}{E - m_{D^0} - m_{D^{\ast 0}} - \frac{\hat{p}^2}{2 \mu_0} + i \varepsilon} \, ,
\end{equation}
\begin{equation}
\label{TT_eq:3_6}
G_3^0(E) = \frac{1}{E - m_{D^+} - m_{D^{\ast -}} - \frac{\hat{p}^2}{2 \mu_+} + i \varepsilon} \, .
\end{equation}

\begin{figure}[!tb]
\centering
\includegraphics[width=0.6\columnwidth]{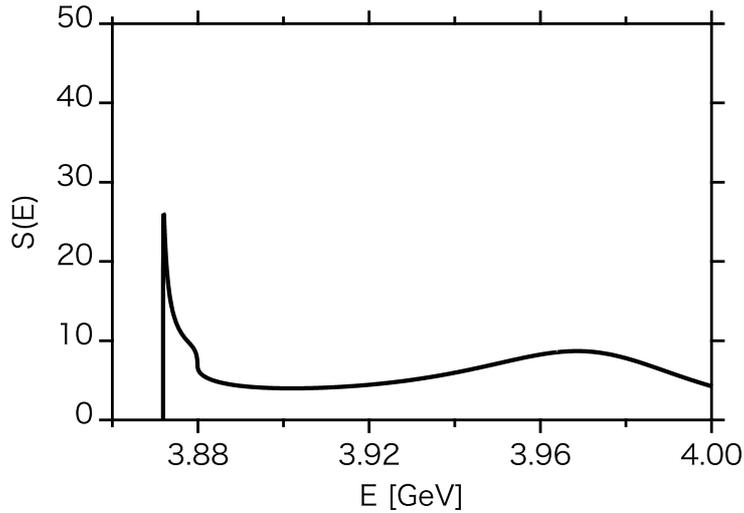}
\caption{The transition strength $S(E)$ with the cutoff $\Lambda = 0.3$ GeV and 
the mass of the $X(3872)$ $m_X = 3.87168$ GeV. 
The $c \bar c \rightarrow X$(3872) strength is 0.051, 
The $S(E)$ is plotted only for the energy above the $D^0 \dsz$ threshold.
The contribution to the bound state 
is not plotted here because it does not have a width.
}
\label{TT_fig:3_1}       
\end{figure}

The calculated transition strength for the cutoff $\Lambda = 0.3$ GeV with the mass of
the $X(3872)$ $m_X = 3.87168$ GeV is shown in Fig.~\ref{TT_fig:3_1}.
The spectrum has a sharp  cusp above the $D^0\dsz$ threshold.
The resonance which corresponds to  the $\chi_{c1}(2P)$ becomes very broad.
The bound $X$(3872) is not plotted in the figure
because it does not have a width in the scheme. 
If we consider the experimental inaccuracy of the
energy and the $X(3872) \to J/\psi \pi \pi$ decay width, 
the bound $X$(3872) peak and the threshold cusp will be merged into
one single peak, which corresponds to the observed $X$(3872) in the $J/\psi \pi \pi$ spectrum.
By integrating $S(E)$ to the $D\dsbar$ continuum state,
one can obtain the transfer strength from the  $c \bar c$-core to 
the bound state.
In this case, the former
 is 0.949 while the latter is 0.051.
The $c \bar c$ core state of the bare mass of 3.950 GeV becomes a resonance state of
$E= (3.974 -{i\over 2}  0.067)$ GeV; 
its peak position is by 24 MeV shifted upward.  

\begin{figure}[!tb]
\centering
\includegraphics[width=0.6\columnwidth]{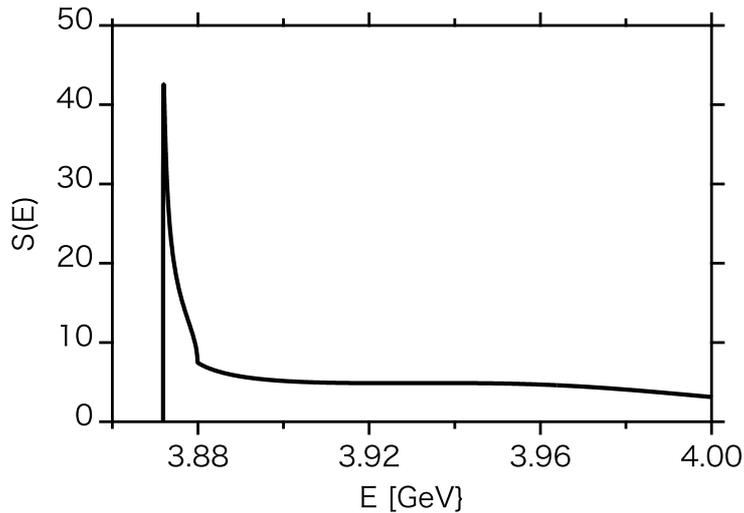}
\caption{The transition strength $S(E)$ with
$\Lambda = 0.5$ GeV and 
 $m_X = 3.87168$ GeV. 
The $c \bar c \rightarrow X$(3872) strength is 0.087.
The legend is as for Fig.\ \ref{TT_fig:3_1}.
}
\label{TT_fig:3_2}       
\end{figure}

In Fig.~\ref{TT_fig:3_2},
we show the transition strength for the cutoff $\Lambda = 0.5$ GeV. 
The spectrum is almost flat at around $E=3.95$ GeV.
In the case of this harder cutoff, the $c \bar c$ core state couples to the $D \dsbar$
continuum of more wider energy range.
As a result, the bump around 3.95 GeV found for the $\Lambda=0.3$ GeV case disappears.
The pole moves to 
$E= (3.971 - {i\over 2} 0.147)$ GeV.
The strength from the $c \bar c$-core to the bound state becomes slightly larger, \i.e.,  0.087.

\begin{figure}[!tb]
\centering
\includegraphics[width=0.6\columnwidth]{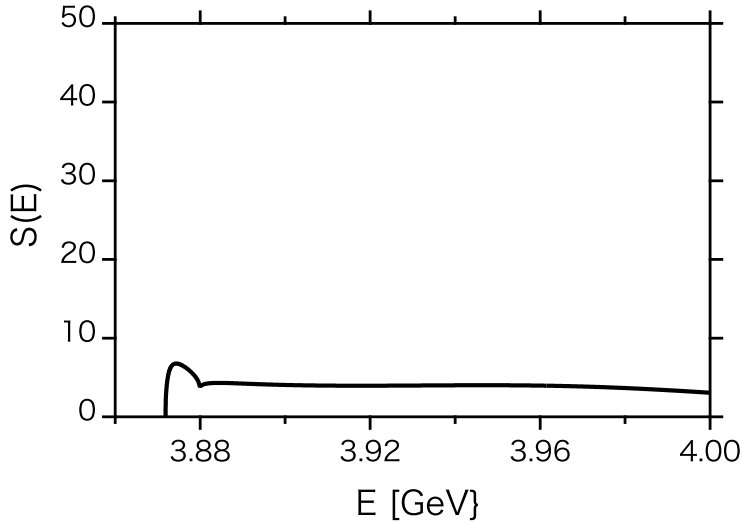}
\caption{The transition strength $S(E)$ with
$\Lambda = 0.5$ GeV and 
 $m_X = 3.8687$ GeV. 
The $c \bar c \rightarrow X$(3872) strength is 0.269.
The legend is as for Fig.\ \ref{TT_fig:3_1}.
}
\label{TT_fig:3_3}       
\end{figure}

Let us now show the effect of the difference in the binding energy. 
In Fig.~\ref{TT_fig:3_3}, we plot the 
transition strength $S(E)$ in the case of the cutoff $\Lambda = 0.5$ GeV and the 
mass of the $X(3872)$ is $m_X = 3.8687$ GeV, \i.e., the more deeply binding case.
The $S$-wave threshold cusp becomes much smaller as the bound state position moves away from the
threshold. The  transfer strength to the bound state in this case is 0.269, much larger than the previous cases.

Since no peak is found around $E=3.95$ GeV 
experimentally, the
$\Lambda = 0.5$ GeV or more is favorable in that sense.
This corresponds to the hadron size $\sim$0.4 fm, 
which is a reasonable value. By setting cutoff of this size,
the shallow bound state and the large decay width for the 
$c \bar c$ peak can be realized simultaneously.
In the following calculation, we use $\Lambda=0.5$ GeV. 
\section{Effect of the interaction between $D$ and $\dsbar$}
\label{DDint}

In this section, we introduce 
the interaction between the $D$ and $\dsbar$ mesons.
We use the Yamaguchi separable potential \cite{YY} for the interaction,
namely,
\begin{equation}
\label{TT_eq:4_1}
 \langle MM' ({\pmb q})| U | MM' ({\pmb p}) \rangle  
   = \frac{-\lambda}{\Lambda^2} \left( \frac{\Lambda^2}{{ q}^2 + \Lambda^2} \right) \, 
  \left( \frac{\Lambda^2}{{ p}^2 + \Lambda^2} \right)
 \,  ,
\end{equation}
where $\Lambda$ is the cutoff, and $\lambda$ is the strength of the interaction.
The Yamaguchi separable potential has been first introduced 
to study the deuteron, the shallow bound state of one proton and one neutron.
So, we consider this interaction is suitable for the present case.
The cutoff $\Lambda$ determines the interaction range and 
therefore, the range is chosen to the typical hadron size here.
For simplicity, 
we take the same value for the cutoff $\Lambda$ in 
Eq.~(\ref{TT_eq:4_1}) as that of Eq.~(\ref{TT_eq:2_5})
in the following calculation.

In order to give a zero-energy bound state only by 
the two-meson interaction, the strength should be
\begin{equation}
\lambda={\Lambda\over \mu_{MM'}}
\end{equation}
with the reduced mass of the system, $\mu_{MM'}$.
For $\Lambda$=0.5 GeV and $\mu_{B\bar B^*}$=2.651 GeV,
this strength becomes 0.1886, which we denote $\lambda_B$ below.
As for the $D\dsbar$, the required strength to have a zero-energy bound state becomes 0.5712.

First let us make a rough estimate of the size of the $D\dsbar$ attraction
using the information from the $B^{(*)}\bar B^*$ system.
Each of the $B\bar B^*$ and  $B^{*}\bar B^*$ systems 
has a $J^P=1^+$ resonance 
by about 2$\sim$3 MeV above the thresholds: $Z_b(10610)$ and $Z_b(10650)$.
Their masses are 10.6072 and 10.6522 GeV, respectively, and 
their mass difference is 45.0 MeV.
The corresponding thresholds, $B\bar B^*$ and $B^*\bar B^*$, are
10.6048 and 10.6504 GeV, respectively, and their energy difference 
 is 45.6 MeV.
This strongly suggests that the two-meson attraction 
is barely strong enough to make a zero-energy bound state (or somewhat weaker), 
and that there are almost no
mixing between the $B\bar B^*$ and $B^*\bar B^*$ $1^+$ states.
The physical origin of the two-meson interaction is probably the light-meson exchange 
and/or the gluonic interaction.
In either case,
the strength of the two-meson interaction for the $D\dsbar$ system
has a similar size to that of the $B\bar B^*$ system,
because the bosons exchanging between the light quarks is considered to 
give the largest contribution.
So, also for the two-meson interaction between 
the $D$ and $\dsbar$, we employ the one with the strength which 
gives a zero-energy bound state for the 
 $B\bar B^*$ systems, $\lambda_B$.

Thus the $D\dsbar$ interaction we employ is:
\begin{equation}
\langle D^0 \dsz ({\pmb q})| U | D^0 \dsz ({\pmb p}) \rangle  = 
  \langle  D^+ \dsm ({\pmb q}) | U | D^+ \dsm ({\pmb p}) \rangle \nonumber\\
 =  \frac{-\lambda}{\Lambda^2} \left( \frac{\Lambda^2}{{ q}^2 + \Lambda^2} \right) \, 
  \left( \frac{\Lambda^2}{{ p}^2 + \Lambda^2} \right)
\end{equation}
with $\lambda = \lambda_B$.
Though, we look into the effects of the $D\dsbar$ attraction
 by changing the value of $\lambda$ from $\lambda_B$.

To use $\lambda_B$ also for the interaction between
$D$ and $\dsbar$ mesons
means that we assume the attraction is  independent of the isospin 
as well as of the heavy quark masses.
Let us make a brief comment why
we do not employ the pion-exchange (OPE) interaction, 
and accordingly a spin-isospin dependent interaction.
The spin-isospin factor of the OPE interaction
between the light quark 
and  the anti-quark is $-(\tau\cdot\tilde\tau)(\sigma\cdot\sigma)$
\cite{Thomas:2008ja}.
\begin{table}[!b]
\caption{The spin-isospin matrix elements of the OPEP by the two-meson states:
$D\dsbar$ $J^{PC}=1^{++}$ and $B^{(*)}\bar B^*$ $I(J^{P})=1(1^+)$.}
\begin{center}
\begin{tabular}{c|ccc}\hline
$ \langle -(\tau\cdot\tilde\tau)(\sigma\cdot\sigma)\rangle$
&$D^0\Dbar^{*0}$
&$D^+D^{*-}$
\\ \hline
$D^0\Dbar^{*0}$
 & 1 & 2  
\\
$D^+D^{*-}$
& 2  & 1 
\\ \hline
\end{tabular}~~~
\begin{tabular}{c|ccc}\hline
$ \langle -(\tau\cdot\tilde\tau)(\sigma\cdot\sigma)\rangle$
&$B^+\bar B^{*0}$
&$B^{*+}\bar B^{*0}$
\\ \hline
$B^+\bar B^{*0}$
&1 & 2
\\
$B^{*+}\bar B^{*0}$ & 2 & 1
\\ \hline
\end{tabular}

\end{center}
\label{tbl:OPEP1}
\end{table}
The factor 
$\langle -(\tau\cdot\tilde\tau)(\sigma\cdot\sigma)\rangle$
becomes $+1$ for the $B\bar B^*$ or the $B^*\bar B^*$ 
diagonal states; {\it i.e.}\ the Yukawa term is repulsive here.
(See Table \ref{tbl:OPEP1}, where we also show those for the $D\dsbar$ systems).
Both of the values corresponds to those
obtained from the heavy meson effective lagrangian\cite{Thomas:2008ja,Ohkoda:2011vj}.
It has been reported that 
 the OPE interaction (with the tensor term and higher order partial wave states) 
makes a bound state\cite{Ohkoda:2011vj}.
There, however, they found that one bound state below the 
$B\bar B^*$ threshold and one resonance above the $B^*\bar B^*$ threshold
rather than two similar resonances.
This occurs because
the factor $\sigma\cdot\sigma$ will also cause the mixing between 
the $B\bar B^*$ and $B^*\bar B^*$ states.
Thus, the spin dependence of OPE interaction seems 
inconsistent with the $B^{(*)}\bar B^*$ experiments,
where the energy difference of the two peaks is almost the same 
as that of the two thresholds.
As was pointed out in Ref.\ \cite{Thomas:2008ja},
the Yukawa term and the $\delta$-function term in the OPE interaction tend to cancel each other.
We consider the OPE interaction is small and the effects of the spin-isospin independent 
attraction are dominant in the present systems. 
\bigskip

Let us go back to the Schr\"odinger equation,
which now includes the two-meson interaction, $U$:
\begin{equation}
\label{TT_eq:4_2}
  \begin{pmatrix}
  m_{c \bar c} - E & V & V \\
  V & m_{D^0} + m_{D^{\ast 0}} + \frac{{\hat p}^2}{2 \mu_0} + U - E & 0 \\
  V & 0  & m_{D^+} + m_{D^{\ast -}} + \frac{{\hat p}^2}{2 \mu_+} + U - E
  \end{pmatrix}
   \,
  \begin{pmatrix}
  c_1 \, | c \bar c \rangle \\
  c_2 \, | D^0 \dsz \rangle \\
  c_3 \, | D^+ \dsm \rangle
  \end{pmatrix} \,  = \,
  \begin{pmatrix}
  0 \\
  0 \\
  0
  \end{pmatrix} \,    
 \, .
\end{equation}
Now the model has one more parameter, $\lambda$, which stands for the coupling strength
of the interaction between the $D$ and $\dsbar$, in addition to 
the cutoff $\Lambda$ and the coupling strength between the $c \bar c$ core and the
two-meson state, $g$. 
The $\lambda = 0$ limit corresponds to the results in Sec.~\ref{form},
where we determined the coupling strength $g$ 
so as to reproduce the observed $X(3872)$ mass
without introducing the direct $D\dsbar$ attraction. 
We call that value $g_0$ in the following and use it as a reference.

In order to study the effects of the interaction between the $D$ and $\dsbar$,
we change the value of the coupling constant $\lambda$. 
For a positive $\lambda$, $g$ should be smaller than $g_0$
in order to  reproduce the observed mass of the $X(3872)$. 
Or, equivalently, when $(g/g_0)^2<1$, 
one has to take $\lambda>0$ to compensate the weakened coupling.
At the $g=0$ limit, the $X(3872)$ becomes a pure $D^0$ $\dsz$ 
hadronic molecular state.
There is no charmonium component nor the $D^+\dsm$ 
component in the $X(3872)$.
There will  be a similar bound state in the $D^+\dsm$ system also,
and the $c\cbar$ core becomes a sharp resonance at around 3.95 GeV.
We consider the actual situation 
is in-between of the two  $\lambda=0$ and $g=0$ limits.
\begin{figure}[!tb]
\centering
\includegraphics[width=0.6\columnwidth]{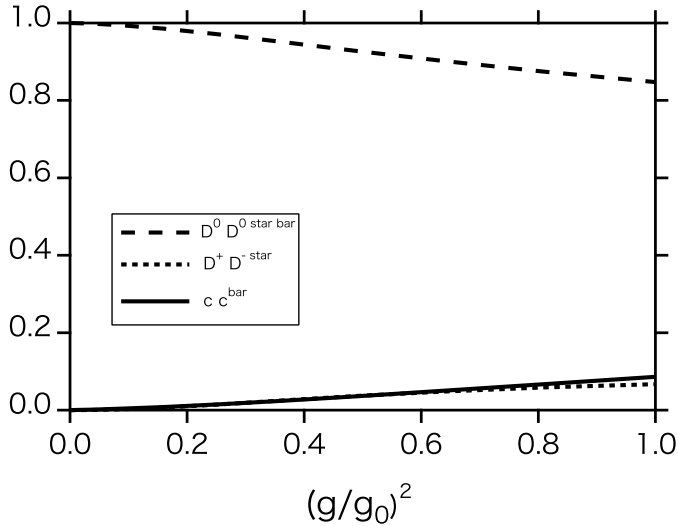}
\caption{Probability of each components in $X$(3872).
We take the mass of the $X(3872)$ $m_X = 3.87168$ GeV with 
the cutoff $\Lambda = 0.5$ GeV.
The solid line shows the size of the $c \cbar$ component in $X(3872)$, 
the dotted line shows that of the $D^+ D^{\ast -}$ and 
the dashed line shows that of the $D^0 \dsz$.
}
\label{TT_fig:4_1}       
\end{figure}

\begin{figure}[!tb]
\centering
\includegraphics[width=0.6\columnwidth]{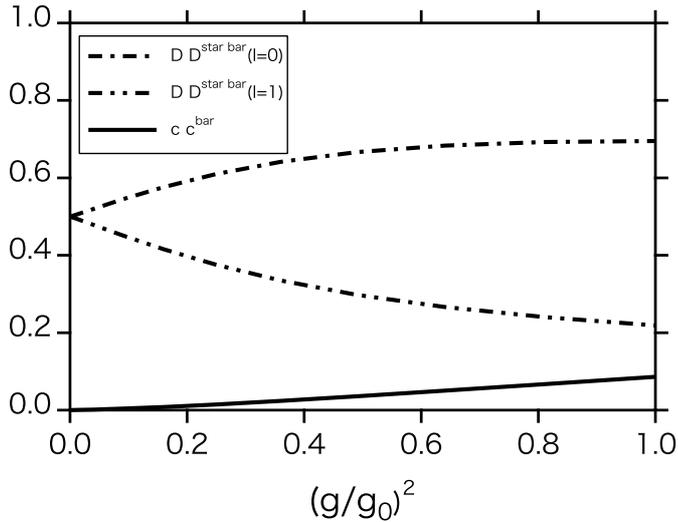}
\caption{Probability of each components in $X$(3872).
The parameters are the same as those in 
Fig.\ \ref{TT_fig:4_1}. 
The solid line shows the size of the $c \cbar$ component in $X(3872)$, 
the dash double dotted line shows that of the isovector $D \dsbar$ and 
the dash dotted line shows that of the isoscalar $D \dsbar$.
}
\label{TT_fig:4_2}       
\end{figure}

 In Fig.~\ref{TT_fig:4_1},
 we show the size of each of the $c \cbar$, 
the $D^0 \dsz$ and the $D^+ \dsm$ components
in the $X(3872)$ wave function in our calculation.
For each values of $(g/g_0)^2$, we re-adjust the value of $\lambda$ to fit the 
 mass of the $X(3872)$ to be 3.87168 GeV.
In Fig.~\ref{TT_fig:4_2}, we also plot the sizes of each of 
the isovector and the isoscalar $D \dsbar$ components.
As the interaction between the $D$ and $\dsbar$ becomes larger (\i.e., $(g/g_0)^2$ becomes smaller), 
the isovector $D \dsbar$ component in the $X(3872)$ wave function becomes larger while 
the isoscalar $D \dsbar$ component reduces to 0.5.

As was mentioned before, experimentally
the isovector component seems to be
about one forth of the isoscalar component (see eq.\ (\ref{TT_eq:1_2})).
Also, the production process of $X(3872)$ suggests that there should be 
a measurable $c\cbar$ component.
From Fig.~\ref{TT_fig:4_2}, one can find 
that these requirements
are fulfilled when $(g/g_0)^2$ is close to  1, namely the $\lambda= 0$ limit.
\begin{figure}[!tb]
\centering
\includegraphics[width=0.6\columnwidth]{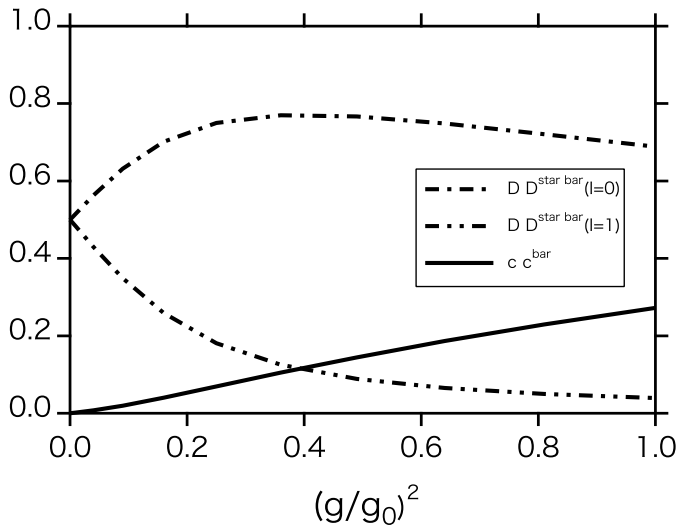}
\caption{Probability of each components in $X$(3872).
We take the mass of the $X(3872)$ $m_X = 3.8687$ GeV with 
the cutoff $\Lambda = 0.5$ GeV.
The legend is as for Fig.\ \ref{TT_fig:4_2}.
}
\label{TT_fig:4_3}       
\end{figure}

When the $D \dsbar$ interaction is switched on, and its strength becomes
$\lambda=\lambda_B$, 
the coupling to the $c\cbar$ core becomes 
$g=0.0427315$, which corresponds to $(g/g_0)^2 = 0.699$.
This point also gives
the appropriate size of the isospin symmetry breaking as well as the measurable $c\cbar$ component.
There each of the components of the $X(3872)$ wave function is: 
\begin{align}
\label{TT_eq:4_3}
 | X \rangle  & = 0.237 \, | c \bar c \rangle - 0.944 \, | D^0 \dsz \rangle 
     - 0228 \, | D^+ \dsm \rangle \notag \\
& = 0.237 \, | c \bar c \rangle - 0.829 \, | D \dsbar ; I=0 \rangle 
     - 0.506 \, | D \dsbar ; I = 1 \rangle 
 \, .
\end{align}
This result means that about 6\% of the $X(3872)$ is the charmonium, about 
69\% is the isoscalar $D \dsbar$ molecule and 26\% is the 
isovector $D \dsbar$ molecule. 
Provided that the rhs of Eq.~(\ref{TT_eq:1_2}) corresponds faithfully
to the ratio of the isovector
to the isoscalar $D \dsbar$ molecular components in the $X(3872)$ wave function as 
it is, the state expressed by Eq.~(\ref{TT_eq:4_3}) is consistent with the experiment.
This situation seems to depend on the $(g/g_0)^2$ value only mildly.

We have also solved the system 
where the mass $m_X = 3.8687$ GeV, namely, by about 3 MeV  more bound case.
The components in such a case are shown in Fig.~\ref{TT_fig:4_3}.
Here, the $c \cbar$ component is much larger than that of $m_X = 3.87168$ GeV.
The size of the $c \cbar$ core component is sensitive to the value of the binding energy.
To make the mass $m_X = 3.8687$ GeV, the strength becomes $g=0.04873$,
which corresponds to $(g/g_0)^2 = 0.750$.
The core component becomes large in this situation, though
the isovector component becomes somewhat smaller. 

\begin{figure}[!tb]
\centering
\includegraphics[width=0.6\columnwidth]{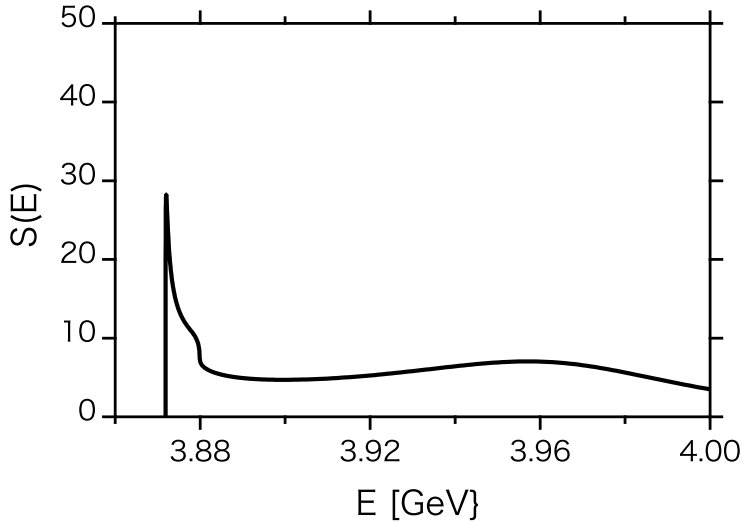}
\caption{The calculated transition strength $S(E)$ with
$\Lambda = 0.5$ GeV and 
 $m_X = 3.87168$ GeV with $(g/g_0)^2 = 0.699$. 
The $c \bar c \rightarrow X$(3872) strength is 0.056.
The legend is as for Fig.\ \ref{TT_fig:2_1}. %
}\label{TT_fig:9}       
\end{figure}

In Fig.\ \ref{TT_fig:9},
we plot the transition strength $S(E)$
for the $\Lambda = 0.5$ GeV and 
 $m_X = 3.87168$ GeV with $(g/g_0)^2 = 0.699$ case. 
Also when the $D\dsbar$ interaction is introduced,
it is found that 
the strength gathers close to the thresholds.
The strength to the $X$(3872) is 0.056.
The peak around the $c\cbar$ core disappears
due to the coupling between the two-meson states and the $c\cbar$ core.
It becomes a resonance of
$E= (3.966 - {i\over 2} 0.091)$ GeV.

Thus, we conclude that in case of the $X(3872)$, 
rather small amount of the interaction is coming from the 
direct interaction between the $D$ and $D^\ast$ mesons 
and that the rest of the attraction is coming from 
the coupling to the $c \cbar$ core state.
Then we have the right size of the isospin symmetry breaking 
as well as  a measurable $c\cbar$ component,
both of which are key features to explain the experiments.
Also, this picture is  consistent with the existence of 
$Z_b$ resonances and absence of the charged $X$.
\section{Application to the other systems}
\label{ExoticHadrons}
\begin{table}[!b]
\caption{The observed mass spectra of $c\bar c$ and $b \bar b$ systems 
with $J^{PC}=J^{++}$ $(J=0,1,2)$ \cite{PDG12} 
and of the quark model with the color-Coulomb, linear confinement, and the color-spin interactions ($Q\bar Q$)
with their lowest $S$-wave 
$D\bar D$ or $B\bar B$ threshold. 
Parameters in the interactions are taken from  \cite{GI85}. All entries are in MeV. }
\begin{center}
\begin{tabular}{cccccccccc}\hline
$c\bar c$&$\chi_{c}(1P)$&$\chi_{c}(2P)$&$\chi_{c}(3P)$&$\chi_{c}(4P)$&
\multicolumn{2}{c}{$S$-wave threshold}
\\\hline
$0^{++}$&3415&-&-&-&$D\bar D$&3730\\
$1^{++}$&3511&-&-&-&$D\bar D^*$&3872\\
$2^{++}$&3556&3927&-&-&$D^*\bar D^*$&4014\\
$c\bar c$($n\,^3\!P_J$)&3526&3964&4325&4642&&\\
\hline
$b\bar b$&$\chi_{b}(1P)$&$\chi_{b}(2P)$&$\chi_{b}(3P)$&$\chi_{b}(4P)$&
\multicolumn{2}{c}{$S$-wave threshold}\\\hline
$0^{++}$&9859&10233&10530&-&$B\bar B$&10559\\
$1^{++}$&9893&10255&10530&-&$B\bar B^*$&10604\\
$2^{++}$&9912&10269&10530&-&$B^*\bar B^*$&10650\\
$b\bar b$($n\,^3\!P_J$)&9884&10252&10543&10791&&\\
\hline
\end{tabular}
\end{center}
\label{Tbl:others}
\end{table}%

In this section, we discuss the possibility to apply the present method to investigate 
 the existence 
of other exotic hadrons as well as the absence of the $Q\bar Q$ states
above the threshold.
If there is a charmonium or a bottomonium state ($Q\bar Q$) above 
the $Q\bar q$ and the $q\bar Q$ meson threshold
and if
the quantum numbers of the system allows $Q\bar Q$ to couple to those two mesons, then this coupling causes 
the effective attraction between the two mesons  
 by the same mechanism as the 
present approach.
Moreover, if the coupling occurs in $S$-wave,
the effective attraction can be larger and the $Q\bar Q$ state 
gains a large decay 
width.

For the overview, we show the observed mass spectra of
$\chi_{cJ}(nP)$ and
$\chi_{bJ}(nP)$
with the quark model results for the $Q \bar Q$ systems
in Table \ref{Tbl:others}
with the lowest $S$-wave threshold of the
$Q\bar q$ and the $q\bar Q$ mesons.
The potential in the
quark model consists of the color-Coulomb, linear confinement, 
and the color-spin interactions. 
The values of the parameters in the interactions are taken from  \cite{GI85}.
Since we have neglected the spin-orbit interaction and the tensor terms,
all of the obtained masses of the $^3P_J$ are the same.
One can see from the table 
that the observed states
below the $S$-wave threshold
roughly correspond to those calculated by the quark model.
Above the threshold, however, 
simple $Q\bar Q$ states are not observed any more.
We argue that they disappear because they have a large width
due to the coupling to the two-meson scattering states.

From the $X$(3872) case, we have learned that 
the $Q\bar Q$ state by about 80 MeV above the 
threshold can contribute to form such an exotic state
assuming that the size of the coupling is similar
to the $X$(3872) case.
Let us check whether
such a state exists in the other systems.

First we discuss the 
$J^{PC} = J^{++}$, ($J=0$, 1, 2) bottomonia, $\chi_{bJ}$.
The observed masses are 
($9859.44 \pm 0.42 \pm 0.31$) MeV
and ($10232.5 \pm 0.4 \pm 0.5$) MeV, 
$\chi_{b0}(1P)$ and $\chi_{b0}(2P)$, respectively,
 ($9892.77 \pm 0.26 \pm 0.31$) MeV and 
($10255.46 \pm 0.22 \pm 0.5$) MeV for the $\chi_{b1}(1P)$ 
and $\chi_{b1}(2P)$, respectively,
and 
 ($9912.2 \pm 0.26 \pm 0.31$) MeV and 
($10268.65 \pm 0.22 \pm 0.5$) MeV for the $\chi_{b2}(1P)$ 
and $\chi_{b2}(2P)$, respectively.
The second radially excited state has been found
at  ($10530 \pm 10$) MeV, and the
observed peak is the mixture of $J = 0, 1, 2$.
The threshold of the $B \bar{B}$ [$B \bar{B^\ast}$] scattering states 
is 10559 [10604] MeV, 
which is by 29 [74] MeV above the $\chi_b(3P)$ mass
and by 232 [187] MeV below the calculated $\chi_b(4P)$ mass.
Since the threshold is much closer to the  $\chi_b(3P)$ than to the  $\chi_b(4P)$,
the effects of the $b\bar b$ states on the 
the $B$ and $\bar B^{(\ast)}$ interaction will probably be repulsive at around the threshold.
As for the the $B^\ast$ and $\bar{B^\ast}$ systems, 
the threshold sits in the middle of the  $\chi_{b1}(3P)$ and $\chi_{b1}(4P)$ states, and the energy differences are about 120-140 MeV.
The $b \bar b$ effects are expected to be small in this case.

We next investigate the $J^{PC} = 0^{++}$ charmonium states. 
The ground state
is $\chi_{c0}(1P)$, and 
its mass is ($3414.75 \pm 0.31$) MeV. 
The $\chi_{c0}(2P)$ state has not
been observed and the theoretical estimation of its mass is 3920 MeV 
\cite{GI85}, whose mass is by 44 MeV lighter than our calculation due to the noncentral force.  
The main $S$-wave decay channel of the $\chi_{c0}(2P)$ state is the 
$D \bar{D}$, whose thresholds is 3730 MeV. 
The $c\bar c$ state is by about 200 MeV above the threshold;
its effects may be attractive, but the size is probably small.

As for the $J^{PC} = 2^{++}$ charmonia, the situation is different from the
$0^{++}$ or $1^{++}$ charmonia. 
 In this channel, 
the first radially excited state, $\chi_{c2}(2P)$,
has been observed, while only the ground states 
have  been observed in the $0^{++}$ and $1^{++}$ channels.  
The reason of this difference is simple in the present picture. 
The $2^{++}$ channel can couple
only to the $D^\ast$  $\bar{D^\ast}$ systems
in $S$-wave; their threshold, 4014 MeV, is rather high
and 87 MeV above the $\chi_{c2}(2P)$ mass.
The calculated mass of the $\chi_{c2}(3P)$ is 4325 MeV,
which is about 300 MeV above the $D^\ast$  $\bar{D^\ast}$ threshold.
So, its effects may be repulsive in this channel.

In summary, the $X$(3872) is found to be surprisingly special.
Although there may be exotic hadrons with the higher partial wave, 
or one has to consider the rearrangement meson channels such as $Q\bar Q$-$q\bar q$
systems, 
 the $c\bar c$ $1^{++}$ channel seems the only promising candidate to form
an $S$-wave exotic hadron by the present mechanism:
a hybrid state of the charmonium and the hadronic molecule.
\section{Conclusion}
\label{conc}
In this work, we have studied the structure of the $X$(3872) as well as
the transfer strength from the $c\cbar$ core to the $D\dsbar$ scattering state.
The system consists of  $D^0\dsz$, $D^+D^{*-}$, and the $2\,{}^3\!P_1$ $c\cbar$ core,
which stands for the $\chi_{c1}(2P)$ if observed.
We have introduced the direct interaction between 
the two mesons,
which is 
just as attractive as the one which makes
a zero-energy bound state  if applied to the $B^{(*)}\bar B^*$ system.
Namely, we assume that this two-meson interaction gives
the $Z_b$(10610) and $Z_b$(10650) resonances.
This interaction, however, is not strong enough 
to make a bound state in the $D\dsbar$ systems alone.
In this model, 
the coupling between the $c \bar c$ core and the $D\dsbar$ two-meson state is also introduced,
 which effectively produces the attraction between the $D$ and $\dsbar$.
We assume that this coupling provides the rest of 
the attraction required to make a bound state in the $D\dsbar$ system, $X$(3872).
Both of the interaction and the coupling are assumed to be isospin independent.
The isospin symmetry breaking in this model solely comes from 
the mass difference between
the neutral and charged $D$ and the $D^\ast$ mesons. 

In the obtained wave function of the $X$(3872), there is about 6\% of the $c \bar c$
core component. 
This size is consistent with a rough estimate from the 
$X$(3872) production rate in the $p \bar p$ collision.
As for the $D \dsbar$ components of the $X$(3872) wave function,
69\% is isoscalar and 26\% is isovector; 
the ratio of the isovector
to the isoscalar $D \dsbar$ components
 is also consistent with the experiments of the final $\pi^2$ to $\pi^3$ decay ratio. 
The present work shows that the structure of 
the $X(3872)$ is not a simple $c \cbar$ nor a simple $D^0 \dsz$ bound state.
It is charmonium-hadronic molecule
hybrid, which is certainly an exotic state. 

Since the $c \bar c$ core cannot couple to the charged $D \dsbar$ states, such as $D^+\dsz$,
the present picture can explain why there exists no charged partners of the $X$(3872). 
Also, it can explain why the $2\,{}^3\!P_1$$c\cbar$ core,
or $\chi_{c1}(2P)$, is not found experimentally 
though it 
has been predicted by the quark model which gives correct mass spectrum below the
open charm threshold;
this core couples strongly to the $D\dsbar$ two-meson state and becomes a resonance
with a very broad width. 

In order to confirm the present picture of the $X(3872)$, 
we consider that the inclusion of the $\rho J/\psi$ and $\omega J/\psi$ channels 
is important because the $X(3872)$ is mainly observed in the 
$X(3872) \to \rho J/\psi \to \pi \pi J/\psi$
and $X(3872) \to \omega J/\psi \to \pi \pi \pi J/\psi$ channels.
We are now performing such calculations and the results will be reported soon.

Recently, Belle Collaboration reported the results of the radiative decays 
of the $X(3872)$ \cite{Belle11}.  They searched the $X(3872) \to \psi' \gamma$ in B decays, but 
no significant signal has been found. On the other hand, $BABAR$ Collaboration has reported 
that ${\mathcal B}(X(3872) \to \psi' \gamma)$ is almost 3 times that of 
${\mathcal B}(X(3872) \to J/\psi \gamma)$ \cite{BaBar09}.
To make the situation clear, it is useful to calculate the radiative decays of 
the $X(3872)$ in the present model including the charmonium structure.
It is left as the future study.
\section*{Acknowledgment}

We would like to thank Professor
K.\ Shimizu and Professor O.\ Morimatsu for
useful discussions.
This work is partly supported by Grants-in-Aid
for scientific research of MonbuKagakushou (20540281) and (21105006).
%
%

%
%
%
\providecommand{\noopsort}[1]{}\providecommand{\singleletter}[1]{#1}%

\end{document}